\documentstyle[prl,epsfig,aps]{revtex} 
\begin{document}  
 
\twocolumn[\hsize\textwidth\columnwidth\hsize\csname@twocolumnfalse\endcsname 
 
\title{Slow relaxations and history dependence of the transport properties  
of layered superconductors} 
\author{Rapha{\"e}l Exartier $^1$ and Leticia F. Cugliandolo
$^{2}$\cite{add1} 
} 
\address{
$^1$ Laboratoire des Milieux D{\'e}sordonn{\'e}s et H{\'e}t{\'e}rog{\`e}nes,  
Tour 13 - case 86, 4 place Jussieu, F-75252 Paris Cedex 05, France 
\\ 
 $^2$ 
\it Laboratoire de Physique Th{\'e}orique de l'Ecole Normale  
Sup{\'e}rieure, 
24 rue Lhomond, 75231 Paris Cedex 05, France and \\ 
 Laboratoire de Physique Th{\'e}orique  et Hautes Energies, Jussieu,  
1er {\'e}tage,  Tour 16, 4 Place Jussieu, 75252 Paris Cedex 05, France 
} 
 
\date\today 
\maketitle 
\begin{abstract} 
We study numerically the time evolution  of  
the transport properties of layered superconductors   
after different preparations.  
We show that, in accordance with recent experiments in  
Bi$_2$Sr$_2$CaCu$_2$O$_8$ performed 
in the second peak region of the phase diagram 
(Portier {\it el al.}, 2001), the relaxation 
strongly depends on the initial conditions and is extremely slow.  
We investigate the dependence on the pinning center density and  
the perturbation applied. We compare the measurements to recent 
findings in tapped granular matter and we interpret our results with a 
rather simple picture. 
\end{abstract} 
\twocolumn  
\vskip .5pc] 
\narrowtext 
 
The behavior of the flux-line array in type II superconductors  
is determined by two competing interactions, the vortex-vortex repulsion  
that favors order in the form of a hexagonal lattice, and the attraction  
between vortices and randomly placed pinning centers due to material  
defects that, as thermal fluctuations, favors disorder.  
This competition generates a very rich static phase diagram~\cite{Blatter}. 
More recently, the evolution of the spatial ordering of vortices 
driven out of equilibrium by, e.g., an 
external current has been analyzed 
analytically \cite{theor_driven}, numerically   
\cite{numer_driven} and experimentally \cite{exp_driven}. 

However, even in the absence of an external force, 
vortex systems show very rich nonequilibrium phenomena.   
Indeed, a hallmark of vortex dynamics is its  
history dependence. A vortex structure prepared by 
zero-field cooling (ZFC), in which  the magnetic field is applied  
right after crossing the transition temperature,  
is expected to be more  
ordered than the one prepared by field cooling (FC), in which the  
magnetic field is applied before going through the transition  
\cite{Kes,Shobu1,Shobu2} (see, however, \cite{Paco}).  
Subsequently, thermal fluctuations  
allow for the rearrangement of the initial configurations. 
Studies of the evolution after FC and ZFC preparations of 
low-$T_c$ and high-$T_c$ type II superconductors 
have been performed with a variety of techniques that we   
non exhaustively summarize below. 
 
The initial current ramp in the IV characteristic of a FC  
2H-NbSe$_2$ sample  
has a large hysteretic critical value, suggesting that 
vortices are strongly pinned \cite{Shobu1,Shobu2}.  
Once the flux lines are depinned, and one applies a subsequent 
ramp, the critical current takes a smaller value.  
In ZFC samples the critical current  
in both ramps is small \cite{Shobu1}. The presence of  
metastability and hysteresis depends on the speed  
of the current ramp imposed to the system \cite{Andrei1}. 
The ac response and   
complex resistivity in 2H-NbSe$_2$ also show that the  
FC state is strongly pinned and disordered while the ZFC state  
is not\cite{Shobu2}. Similar effects have been exhibited in high  
$T_c$ supercondutors. For instance, Josephson plasma resonance experiments in 
Bi$_2$Sr$_2$CaCu$_2$O$_8$ (BSCCO) 
suggest that the resonant field in FC samples is stationary 
while in ZFC samples it approaches the FC value asymptotically~\cite{Matsuda}. 
Still, thermomagnetic history effects in the solid  
vortex phase of pure twinned  
YBa$_2$Cu$_3$O$_7$ single crystals have been recently observed in  
ac susceptibility measurements \cite{Sergio}. On the numerical side, 
Olson {\it et al}  
studied how the fixed drive 
induced velocity (voltage response) depends upon time after different  
sample preparations, close to the order (3D) - disorder (2D)  
transition \cite{Zim}, focusing on the effect of  
superheating or supercooling the ordered and disordered phases \cite{Zel}.   
  
In this Letter we concentrate on a recent study of the long-time  
transport properties of  
BSCCO monocrystals after different preparations 
\cite{Portier}.  
The experimental protocol is as follows. 
The same working conditions, given by a  
temperature $T=4.5K$ and a magnetic field, $\mu _o H=1.5T$, 
perpendicular  
to the $c$-axis  
are initially attained via a FC or a ZFC procedure.  
These values fall in the second peak region \cite{Shobu1}. 
After a waiting-time  ${t_w}=30\mbox{min}$  
a triangular pulse of current of duration $\tau_p\sim 10\mu s -500\mu s$ 
is applied and an IV characteristic is recorded. 
During a second waiting period of the same duration  
${t_w}$ the sample is let freely evolve driven only by thermal fluctuations. 
The same external current is then applied and a second IV characteristic  
is recorded. This procedure is repeated so on and so forth.  
The first observation is that the shape of the IV 
loop changes as time elapses. In order to quantify its change,  
Portier {\it et al} chose a threshold tension, $V_{\sc th}$, 
defined the threshold current, $I_{\sc th}$, as 
the corresponding intensity in the IV characteristics, and monitored 
the dependence of $I_{\sc th}$ on the discrete  times $t_n=n (t_w+\tau_p)$,  
$n=0,1,\dots$. 
$I_{\sc th}$ slowly relaxes in time, in a way that  
depends on the history of  
the sample. In the FC sample, after a seemingly static period 
that lasts until $t\approx 10^4\, s$, $I_{\sc th}$ relaxes $\approx
25\%$ of its
initial value over one decade in time, with a logarithmic decay.   
In the ZFC sample instead, $I_{\sc th}$ smoothly increases
by less than a tenth of its initial value over $6$ decades.  
The asymptotic value of $I_{\sc th}$ is the same for both  
cooling procedures. The samples  
evolves in very long time scales, of the order of days.   

The microscopic interpretation  
of experiments that probe the flux motion via the measurement of 
transport properties is not always straightforward.  
We reproduced the latter experiment with numerical  
simulations using a model of pancake vortices in a layered superconductor 
that successfully captures many  
of the observed effects in type II superconductors  
\cite{Zim,Bariloche,PhaseDiag,Log,Olson,Olson-Vin}. 
For adequately chosen sets of parameters our 
 results  are in qualitative accord with the  
measurements of Portier {\it et al} \cite{Portier}. 
The advantage of  
using numerical simulations is twofold. First, 
it allows us to  
simply explore the effect of the microscopic parameters in the model.  
Second, it allows us to grasp what vortices 
are actually doing by direct visualization.

The model takes into account the long-range  
magnetic interactions between all pancakes.
Since a flux line is essentially massless \cite{Blatter}  
we use overdamped Langevin dynamics. 
The equation of motion for a pancake located at a position  
${\bf R}_i=({\bf r}_i,z_i)=(x_i,y_i,z_i)$, (${\bf z}\equiv \hat c$), is 
\begin{equation} 
\eta \frac{d{\bf r}_i}{dt}=\sum_{j\neq i}{\bf F_v}(r_{ij},z_{ij}) 
+ 
\sum_p{\bf F_p}(r_{ip})+{\bf F} 
\; , 
\label{eq:Lang} 
\end{equation} 
where $r_{ij}=|{\bf r}_i-{\bf r}_j|$ and  
$z_{ij}=|z_i-z_j|$ are the in-plane and  
inter-plane distances between pancakes $i$ and $j$. 
$\eta$ is the friction coefficient.
The driving force per unit length due to an in-plane  
current {\bf J} acting on a vortex  is  
${\bf F}=\Phi_0\, {\bf J}\times \hat{\bf z}$, with $\Phi_0$ the quantum of
magnetic flux.  
Quenched random point disorder is modeled by pinning centers that occupy  
uncorrelated random positions, taken from a uniform distribution,  
on each layer.  
They exert an attractive force on the vortices,  
\begin{equation} 
{\bf F_p}(r_{ip})=-\frac{2A_p}{a_p^2}\; e^{-(r_{ip}/a_p)^2} {\bf r}_{ip} 
\; , 
\end{equation} 
where $A_p$ measures the amplitude of this force,  
$a_p$ is the pinning range and $r_{ip}=|{\bf r}_i-{\bf r}_p|$  
is the in-plane distance between vortex $i$ and a pinning site at  
${\bf R_p}=({\bf r}_p,z_i)$. 
The magnetic interaction between pancakes is  
\begin{eqnarray} 
{\bf F_v}(r,0)&=&\frac{A_v}{r} 
\left[1-\frac{\lambda_{||}}{\Lambda}\left (1-e^{-r/\lambda_{||}} 
\right ) \right ] \label{Fv(r,0)}  
\; ,  
\\ 
{\bf F_v}(r,z)&=& 
-s\, \frac{\lambda_{||}}{\Lambda}\frac{A_v}{r}\left [ e^{-|z|/\lambda_{||}} 
-e^{-R/\lambda_{||}} \right ] \label {Fv(r,z)} 
\; ,  
\end{eqnarray} 
with $R=\sqrt{z^2+r^2}$, 
$\Lambda=2\lambda_{||}^2/d$ the 2D thin-film  screening length,
$d$ the inter-layer spacing and $\lambda_{||}$ the in-plane penetration 
length.
One can vary the relative strength of the interlayer coupling by tuning   
the prefactor $s$ \cite{Zim,PhaseDiag,Olson}.  
The model in Eqs.~(\ref{Fv(r,0)}) and (\ref{Fv(r,z)}) is valid  
in the limit  $d\ll \lambda_{||}\ll \Lambda$~\cite{Clem}. 
We normalize length scales by $\lambda_{||}$, energy scales by  
$A_v=\Phi_0^2/4\pi^2\Lambda$ and time by $\tau=\eta\lambda_{||}^2/A_v$, that 
for BSCCO yields $\tau\sim 1 \mu s$. 
We consider $N_v$ vortices and $N_p$  pinning centers per layer in $N_l$  
rectangular layers of size $L_x\times L_y$ ($L_x=\sqrt{3}/2L_y$). 
The normalized vortex density is  $n_v=B/\Phi_0=N_v/(L_x L_y)$. 
Following \cite{Bariloche} we consider $n_v=0.12$, $A_p/A_v=0.2$,  $a_p=0.1$, $d=0.01$
and $\Lambda=200$ which correspond to BSCCO \cite{Clem}. 
We use $N_v=25$ vortices in up to $N_l=4$ layers. 
The value of $a_p$ as compared to $n_p$ is of first importance and will 
be discussed later on. We use periodic boundary condition in all  
directions and the periodic  
long-range in-plane interaction is evaluated with an exact and  
fast converging sum \cite{Log}. 
All the simulations are at $T=0$,  
with a time-step $0.01$, and averaged over $90$ realizations. 
  
We mimic the external current applied 
with an increasing (in steps)  
force ${\bf I}$ parallel to the $x$ axis. 
It results in an average motion of the pancakes, with velocity   
$V=[\langle V_x(t)\rangle]=(N_vN_l)^{-1}  \sum_{i}  
[\langle dx_{i}/dt \rangle]$. 
(The angular and square brackets represent an average over different realizations of the  
dynamics and disorder, respectively.) 
We match the notations in Ref.~\cite{Portier} by  calling $I_{\sc th}$  
the value  
of the applied force which leads to an average speed equal to a previously chosen  
threshold $V_{\sc th}$. 
We have checked that the results do  
not depend qualitatively on the precise value of $V_{\sc th}$, if {\it e.g.} 
$V_{\sc th} \leq 0.2$ for $I_{\sc max}=1.4$ and $n_p=0.25$. 
 
\begin{figure}[7] 
\centerline{ 
\epsfxsize=1.7in\epsffile{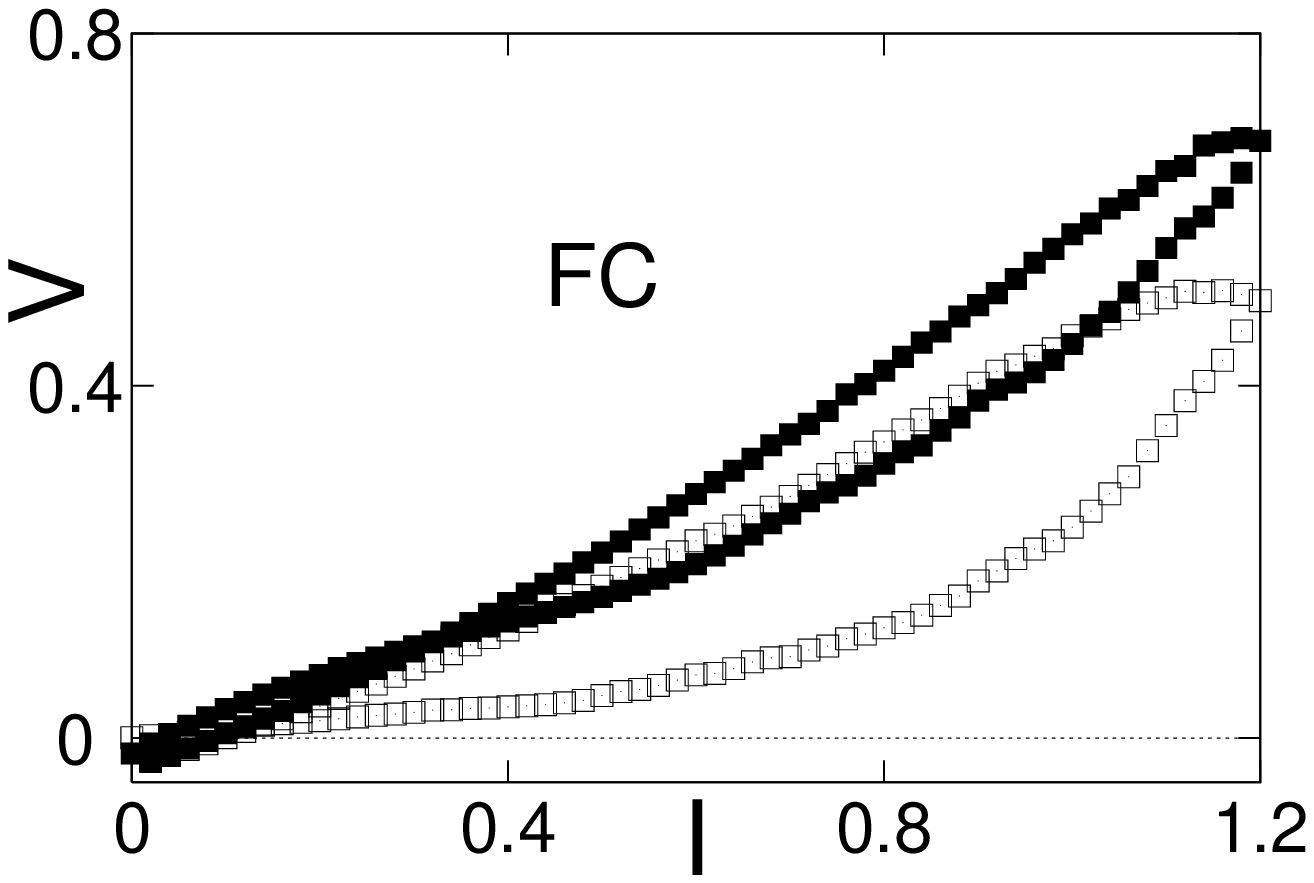} 
\epsfxsize=1.7in\epsffile{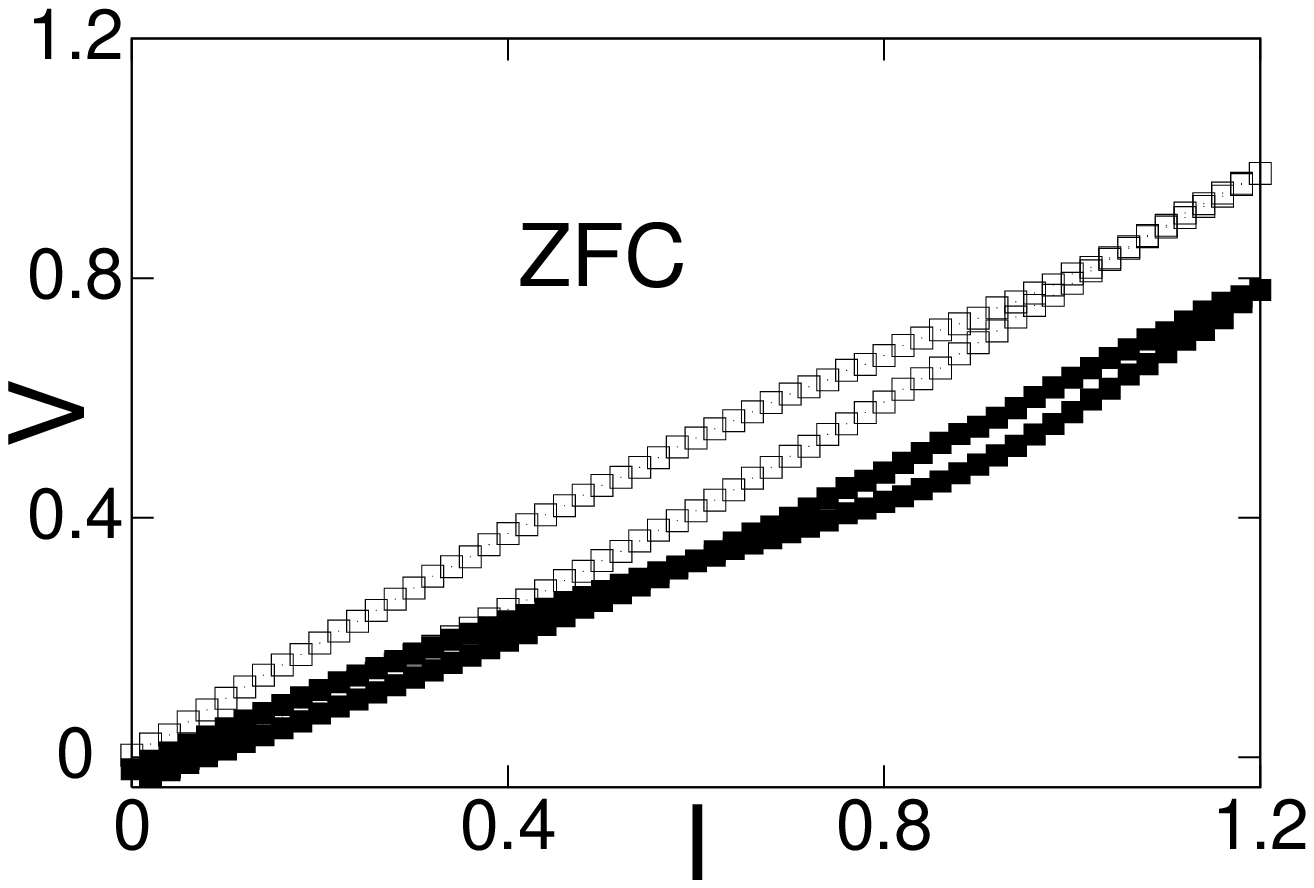} \\}  
\vspace{0.25cm} 
\caption{Current-voltage curves at times $t_1=0$ (open circles)  
and $t_2=20$ (filled circles) 
for a monolayer of pancake vortices  
with $n_p=0.12$, $t_w=6$,  
$\tau_p=14$ and $I_{\sc max}=1.2$. 
} 
\label{F2.2} 
\end{figure} 
 
Experimentally, vortices are created on the surface of the sample when the  
magnetic field is turned on. It happens during cooling for FC samples and later for ZFC samples.  
Afterwards, they penetrate the sample and new ones are created on the 
border.  This procedure  
is difficult to implement in this model.  
We differentiate the FC and ZFC initial states by choosing different  
initial repartitions of vortices. For the FC samples,  
we choose an initial condition with each vortex  
pinned to a defect, in which the distance to the  
nearest pinning center is less than the pinning range $a_p$.  
 To reproduce the ZFC samples, we choose an 
ordered hexagonal lattice \cite{Kes,Shobu1,Shobu2}.
This procedure neglects the critical 
state and it is expected to describe the behavior in the bulk. 
 
In Fig.~\ref{F2.2} we display the first two  
IV loops for a 2D system that starts  
from a pinned (FC) or hexagonal (ZFC) 
initial condition.  
As expected~\cite{Andrei1,Olson-Vin},  
the first ramp  shows hysteresis and a large critical current 
for the FC sample. 
In the second ramp the loop moves and it is partially closed.  
The ZFC sample has a much lower critical current  
(that may even vanish). 
The pulse we used is as fast as the fastest used experimentally,
$\tau_p=14$ in real time; consequently, 
the critical currents are very weak. Moreover, since the waiting-time  
between two successive pulses is here $t_w \sim \tau_p$, 
while in \cite{Portier} $t_w \gg \tau_p$, we are perturbing the system  
in a much stronger manner than done experimentally. By the end of this  
Letter we shall study the dynamics under a less invasive probe.

\begin{figure}[7] 
\centerline{ 
\epsfxsize=1.7in\epsffile{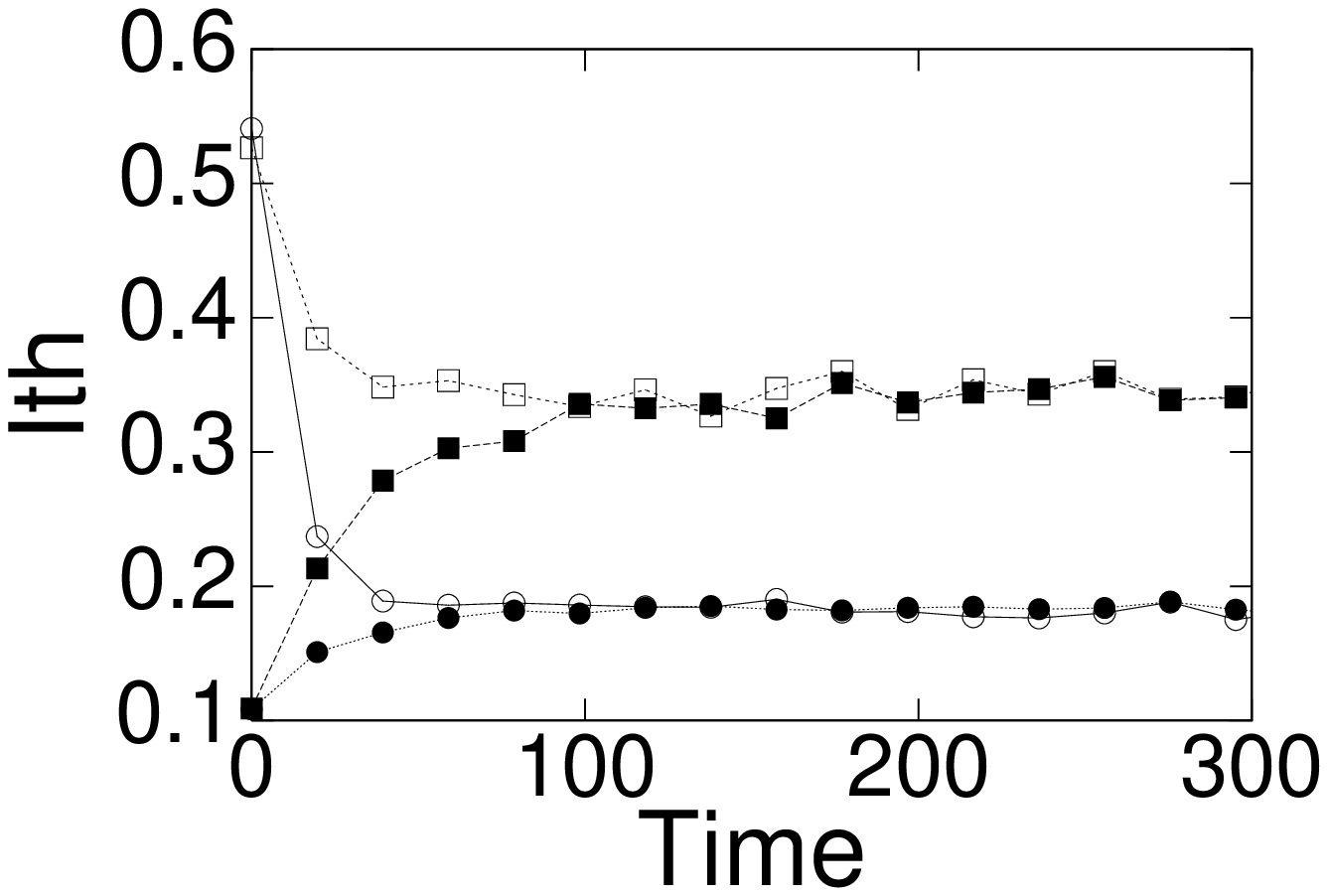} 
\epsfxsize=1.7in\epsffile{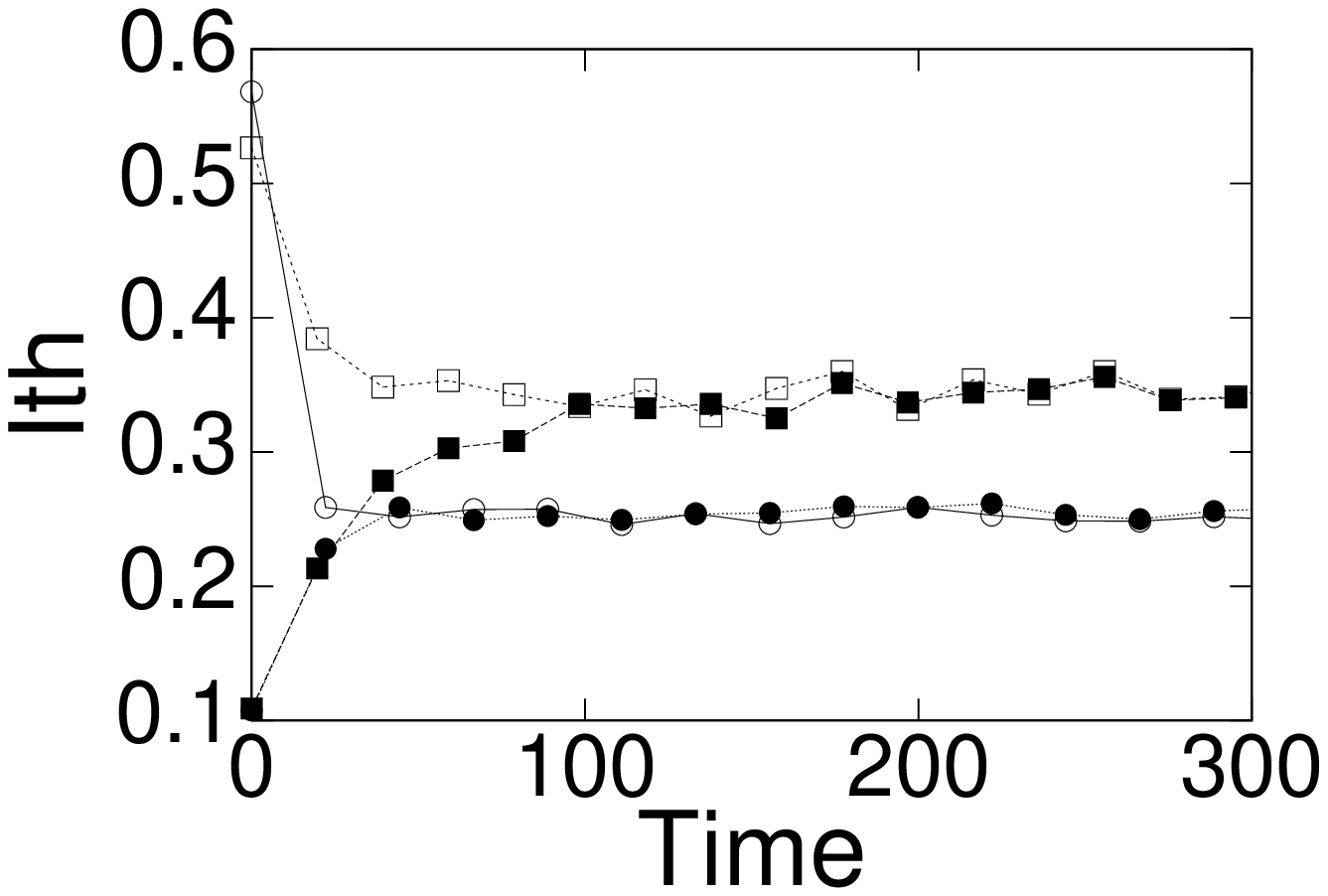}  
}  
\vspace{0.25cm} 
\caption{ 
Left: The threshold current $I_{\sc th}(t_n)$ against time  
in a monolayer
with $n_p=0.12, \, 0.25$ (upper to lower asymptotic value).  
Open symbols: pinned, FC; filled symbols: lattice, ZFC. $I_{\sc max}=1.4$. 
Right: $I_{\sc th}(t_n)$  for $I_{\sc max}=1.4$,  
$\tau_p=14$, $t_w=6$ (above) and  
$I_{\sc max}=1.6$, $\tau_p=26$, $t_w=4$ (below). 
FC (open symbols) and ZFC (filled symbols). 
$n_p=0.25$. 
} 
\label{FC-ZFC} 
\end{figure}

The time-dependence of the IV loop is quantified in Fig.~\ref{FC-ZFC}-left 
where we show the evolution of $I_{\sc th}$ 
 against $t_n$ for two $n_p$. The curves  
correspond to pinned (open circles) and  hexagonal (filled squares)
initial conditions. The first two points  
in the lower curves  
for the pinned and hexagonal initial conditions correspond to the  
IV curves in Fig.~\ref{F2.2}. 
There is  an initial fast motion 
that translates into a rather fast drop in $I^{\sc fc}_{\sc th}$ 
and a fast increase of $I^{\sc zfc}_{\sc th}$. Later,  
the curves approach their common asymptote in a slower than exponential  
manner. The qualitative behavior of these curves  
resembles the ones in \cite{Portier} though with 
much shorter relaxation times. 
  
A closer inspection of the model and procedure suggests that  
the precise shape of these  curves may depend on several  
parameters like $n_p$, the form of the ramp and 
the choice of the sequence $t_n$.  
For very small values of  $n_p$ 
no interesting effect is expected since there are no  
enough pinning centers to stop the  vortices. 
For intermediate densities, {\it e.g.} $n_p=0.12$,  
we found the curves in Fig.~\ref{FC-ZFC}-left.
The asymptotic value of $I_{\sc th}(t_n)$ increases with 
$n_p$.
Eventually, if $n_p$ is so large that the distance between  
pinning centers is smaller than the range of the 
pinning force, \textit{i.e.} $n_p>\pi^{-1}(2a_p)^{-2}$, there is no  
difference between ZFC and FC preparations and after a very few 
iterations the two curves collapse.

The maximum intensity in the ramp, $I_{\sc max}$,  
is also an important parameter. 
The asymptotic value of $I_{\sc th}(t_n)$  
decreases with $I_{\sc max}$. 
For a  stronger $I_{\sc max}$, e.g.
$I_{\sc max}=1.6$ in Fig.~\ref{FC-ZFC}-right,  
the memory of the initial condition is very quickly lost since the  
vortices are pushed too strongly by the force.  
The perturbation itself, {\it i.e.} the sequence 
of current/force pulses, is driving the system to different steady  
states characterized by $I_{\sc th}(t_n \to\infty)$.
 
\begin{figure}[h] 
\centerline{ 
\epsfxsize=1.7in\epsffile{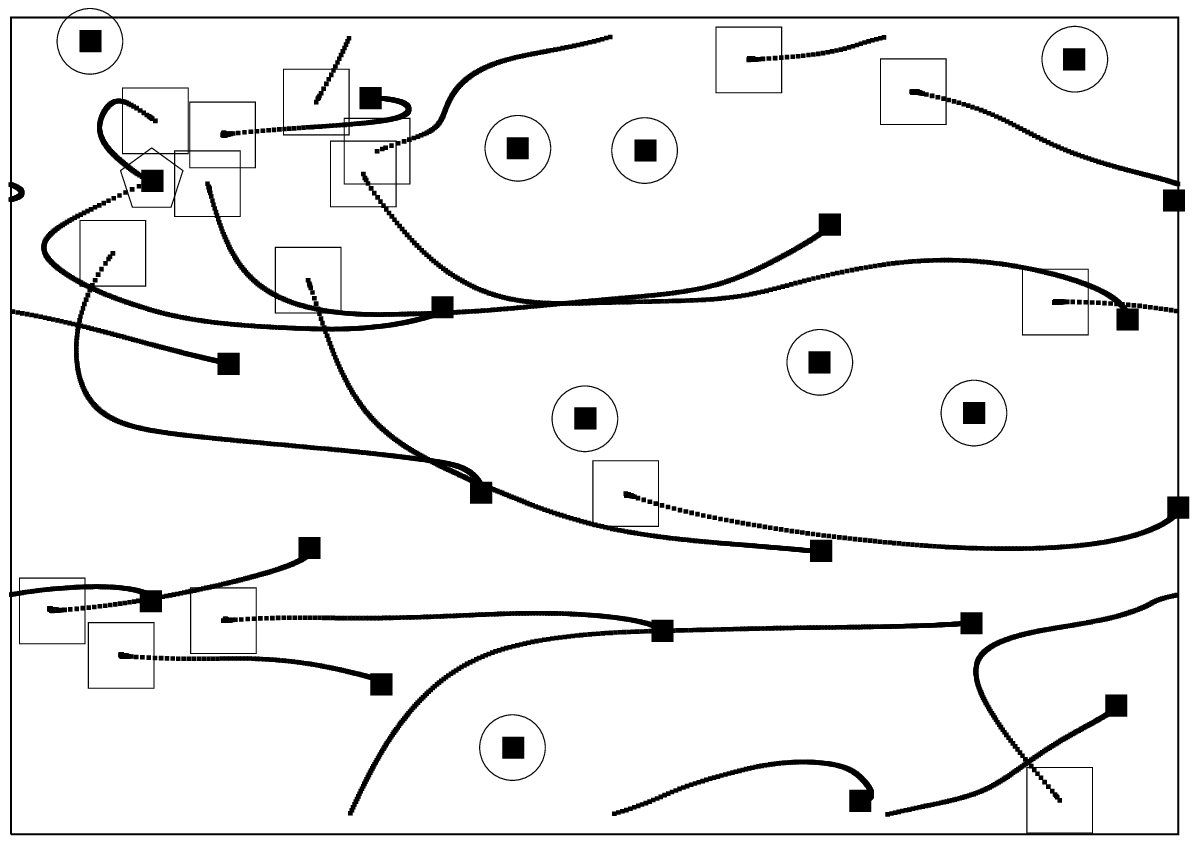} 
\epsfxsize=1.7in\epsffile{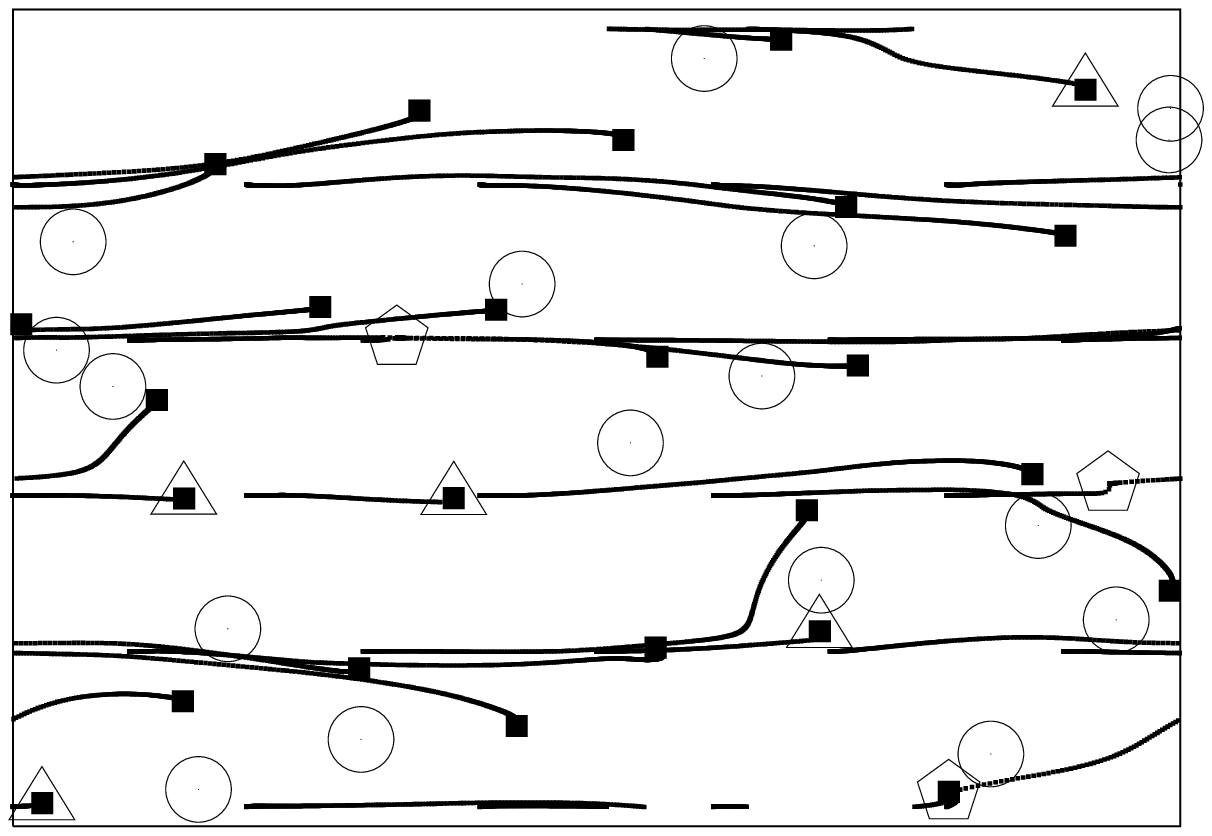}  
\\}  
\vspace{0.25cm} 
\caption{Vortex trajectories of the FC (left) and ZFC (right)  
samples during the first pulse (lines). Pinning centers where a vortex
gets pinned are represented with triangles, those from which  a vortex
depins are represented with squares and those where vortices 
get pinned and later depin are represented with pentagons. Pinning centers
unaffected by the pulse (whether there is a vortex permanently 
pinned to them or not) are represented with  circles. 
} 
\label{photos} 
\end{figure} 
 
Analyzing the motion of vortices during each pulse 
helps us understanding these results. During the first pulse 
a fraction of vortices in the FC sample remains pinned 
while the rest rapidly depin and move in the sample. 
How many of them depin depends on $\tau_p$ and $I_{\sc max}$. 
On the left of Fig.~\ref{photos} we display 
the vortex trajectories  with lines and the positions of  
pinning centers with symbols; 15 vortices move  while the rest remain pinned. 
Some trajectories join two pinning centers. 
During the quiescent time $t_w$ a number of vortices  
may get again pinned, depending on $n_p$, the strength of the  
different interactions and thermal fluctuations.
In the second pulse new vortices depin and, consequently,  
$I_{\sc th}$  decreases.

Instead, 
the initially ordered sample rapidly reaches a smectic flow  
\cite{theor_driven} (Fig.~\ref{photos}-right). 
During this moving period, the vortices that move close to  
pinning centers get pinned;
other vortices are pulled by the force and depin, but 
get later pinned by another pinning center. 
By the end of the first ramp, the hexagonal lattice has been deformed 
and 6 vortices are pinned.  
At $T=0$  the displacement 
during the quiescent time $t_w$ is rather small. 
This mechanism repeats in 
subsequent periods and the threshold intensity increases in time.  
This picture is confirmed by the study of the position of each 
vortex relative to its closest pinning center as a function of $t_n$.  
In the initially pinned sample, on average and in time, 
vortices move away from the pinning centers. 
In an initially depinned
sample, vortices get pinned by
a pinning center. A Delaunay triangulation analysis 
shows that at long times, e.g.
$t_n=1500$, both initially ordered and 
pinned systems become a deformed hexagonal lattice.


We wish to check if this is a plausible explanation of the  
results in~\cite{Portier}.  
Certainly, we have almost continuously stirred the sample 
while the probing pulses applied experimentally, 
being  widely separated in time, are much weaker. 
Unfortunately we cannot match the experimental conditions 
$t_w\gg \tau_p$ ($t_w=30\mbox{min}\equiv 10^9$ iterations!). On the other hand,
mechanical vibrations have not been totally eliminated from the 
experiment and might induce an important supplementary 
means for relaxation~\cite{Portier}. 

\begin{figure}[7]
 \vspace{0.28cm}
\centerline{ 
\epsfxsize=2.7in\epsffile{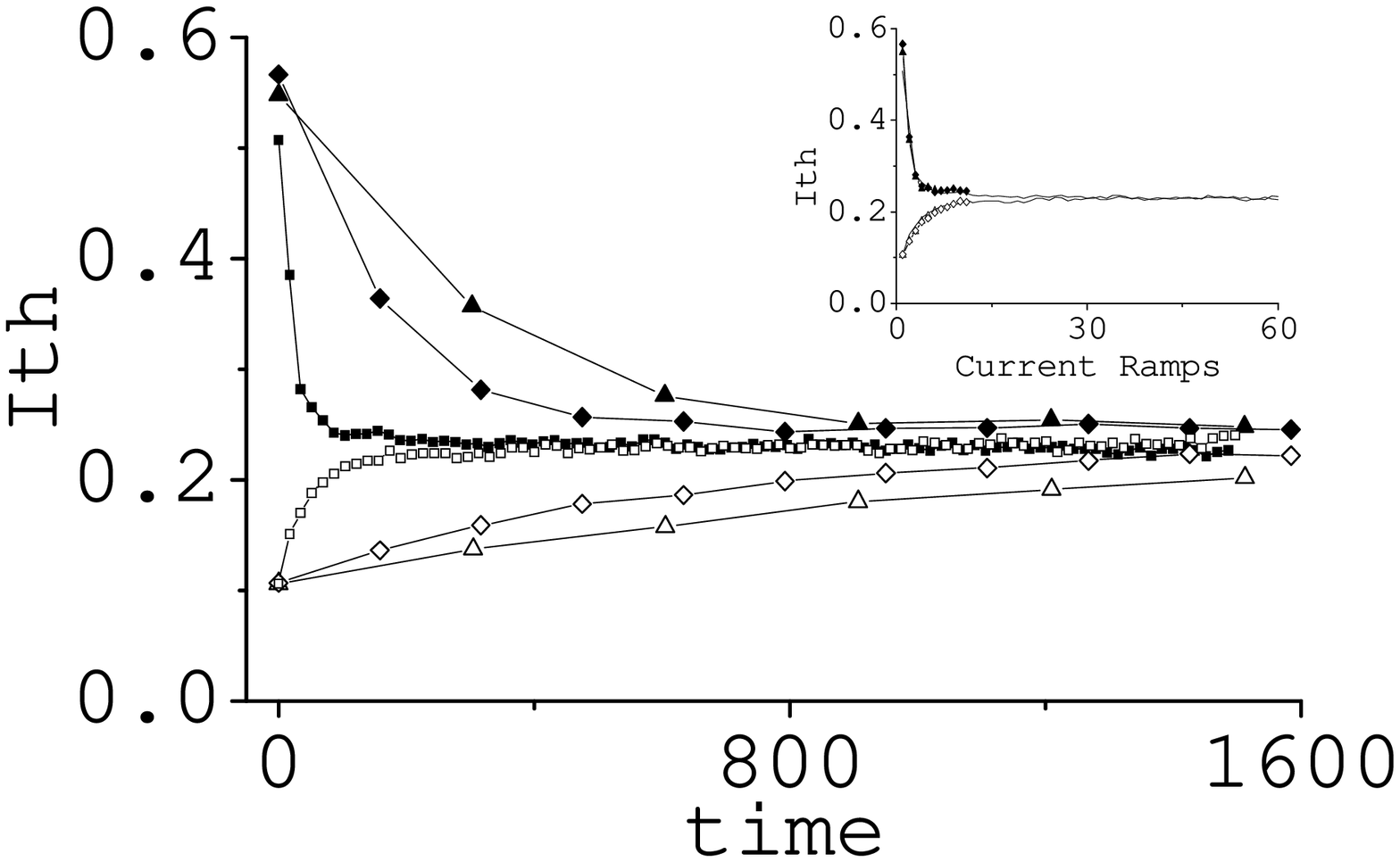}  
} 
\vspace{0.2cm} 
\caption{The threshold intensity against time for FC (filled symbols) 
and ZFC (open symbols) preparations and  $t_w=3,144, 288$, $\tau_p=14.4$
(circles, diamonds, triangles). In the inset, the same curves are plotted
in function of the number of current ramps applied and they rescale perfectly.  
} 
\label{tw_dep} 
\end{figure}

Another important difference between our numerical study and 
the experiment in~\cite{Portier} is that we have worked at $T=0$ (with the aim 
of reducing the thermal noise and improving the averaging) 
while experiments are done at low but non-vanishing $T$. 
The $T=0$ protocol is very close to the ones used to 
probe the dynamics of granular matter. Indeed, 
temperature is totally irrelevant in these systems 
and the rearrangement of grains is induced by pumping
energy in the sample in, e.g., the form of periodic taps~\cite{Sid}. 
If the vortex system at $T=0$
behaves as granular matter, all the dynamics must then take place 
during the probing pulses while the systems should be essentially 
static in a  metastable configuration during $t_w$. 
In Fig.~\ref{tw_dep} we analyse the dependence of the relaxation
time-scale on $t_w$ 
by comparing $I_{\sc th}^{\sc fc,zfc}(t_n)$ for  
$t_w=3, 144, 288$ and $\tau_p=14.4$ in all cases.  
The longer $t_w$ the slower the approach to the asymptotic value.   
The inset shows $I_{\sc th}$ against the number of  
perturbing cycles $n$. The three curves now collapse showing that at 
$T=0$ the evolution of $I_{\sc th}$ takes place during $\tau_p$ and it is  
almost completely due to the perturbation. It would be very
interesting to check if this also happens experimentally.

The mechanism for the very slow relaxation 
in the experimental system may be the one put forward in the previous 
paragraph. The relaxation of the FC and ZFC
samples are controlled by the parameters in the model ($n_p$,
$I_{\sc max}$, etc). Hence, by changing the values of the parameters, 
one can explain the difference in the experimental observations
described in the introduction.
  We have also
simulated 3D systems with different  number of  
layers (1 to 4) and interlayer coupling strengths $s$ (1 to 9),  
and observed no qualitative difference in the  
$I_{\sc th}$ relaxation  
though its time-scale increases considerably.

We wish to thank very useful discussions with P. Chauve, F. De la
Cruz, G. d'Anna, D. Dominguez, 
T. Giamarchi, A. Kolton, P. Le Doussal, S. Valenzuela  
and especially F. Portier and F. Williams  
for communicating their results to us prior to publication, and  
continuous discussions during this work.

\end{document}